# A Little About Folklore

E.G. Maksimov[1], O.V. Dolgov[2]

[1]*P.N. Lebedev Physical Institute, Russian Academy of Sciences, 119991 Moscow, Russia*
[2]*Max-Planck-Institut für Festkörperforschung, D-70569 Stuttgart, Germany*

Comment on paper [1] by Anderson is presented. This Anderson's work is shown to contain a number of inaccurate and ungrounded statements. We point out, in particular, that the total static dielectric function not only can be negative, but inevitably has a minus sign in many stable systems. We also demonstrate that in many metals, the effective electron-phonon interaction is stronger than the direct Coulomb repulsion, without taking into account the ladder-sum renormalization or pseudization of the Coulomb repulsion. Other issues touched in paper [1] are also discussed.

Recently, P.W. Anderson has published the paper "Do we need (or want) a bosonic glue to pair electrons in high-$T_c$ superconductors?" [1], which, in our opinion, contains a number of inaccurate and ungrounded statements. One of these statements is related to the old discussion between P.W. Anderson and group of the Lebedev Physical Institute. It concerns possible signs of the static dielectric constant. Previously, Cohen and Anderson in their early paper [2] used the simple expression for the electron-electron interaction in a metal:

$$V(\mathbf{q},\omega) = \frac{4\pi e^2}{q^2 \varepsilon_{\text{tot}}(\mathbf{q},\omega)} \quad , \tag{1}$$

where $\varepsilon_{\text{tot}}(\mathbf{q}, \omega)$ is the momentum- and frequency-dependent total dielectric function. This function includes the screening due to the Coulomb interaction and a contribution of the electron-phonon interaction (EPI). By considering that the static dielectric function $\varepsilon(\mathbf{q}, 0)$ satisfies the inequality

$$\varepsilon(\mathbf{q},\omega) > 0 \quad , \tag{2}$$

the authors of Ref. 2 have shown that this inequality results in a strong limitation on possible values of the critical temperature of superconducting transition $T_c$. This limitation is due to the interrelation between the Coulomb repulsion constant $\mu$ and the electron-phonon coupling constant $\lambda$, which follows from inequality (2):

$$\mu > \lambda \quad . \tag{3}$$

This means that an effective electron-electron interaction is repulsive. Superconductivity cannot occur in such a system. As Tolmachev has shown [3], there is a the ladder-sum renormalization or pseudization of the Coulomb repulsion to a smaller value,

$$\mu^* = \frac{\mu}{1 + \mu \ln(\varepsilon_F / \omega_D)} \quad , \tag{4}$$

where $\varepsilon_F$, is the Fermi energy and $\omega_D$ is an average phonon energy. Thus, the effective interaction can become attractive, $\lambda - \mu^* > 0$. The term $\lambda - \mu^*$ appears in various approximate expressions for $T_c$ and at $\lambda = \mu$ the limitation on maximum value of $T_c$, that follows from those expressions, leads to the inequality

$$T_c \lesssim \varepsilon_F \exp(-1/\lambda) \lesssim 10 \text{ K (!)}$$

The above arguments are mostly repeated in paper [1].

We discussed this problem in detail in our review [4]. Here, in order to solve at least this issue once and for all, we briefly repeat our results that demonstrate the existence of negative values of $\varepsilon(\mathbf{q}, 0)$ in many simple systems just in their stability region. One of simple systems with $\varepsilon(\mathbf{q}, 0)$ negative at any momentum $\mathbf{q}$ is the Wigner crystal. The dielectric function $\varepsilon(\mathbf{q}, \omega)$ of classical Wigner crystal has been thoroughly considered by Bagchi [5] who showed that

$$\frac{1}{\varepsilon(\mathbf{q},\omega)} = 1 - \frac{\omega_{pl}^2}{q^2} \sum_\lambda \frac{(\mathbf{q}\mathbf{e}_{\mathbf{q}\lambda})^2}{\omega^2(\mathbf{q},\lambda) - \omega^2} \quad . \tag{5}$$

Here, $\omega_{pl}$ is the plasma frequency of charged particles:

$$\omega_{pl}^2 = \frac{4\pi n e^2}{m} \quad , \tag{6}$$

where $m$ and $n$ are the mass and density of the particles, respectively, $\mathbf{e}_{\mathbf{q}\lambda}$ is the polarization vector of phonons, and $\omega(\mathbf{q}, \lambda)$ is the phonon frequency.

In a cubic Wigner crystal with one atom per unit cell, there are three phonon modes, two transversal modes with acoustic dispersions at small $\mathbf{q}$ vectors and one longitudinal mode whose frequency $\omega(\mathbf{q}, \lambda)$ approaches to $\omega_{pl}$ at $q \to 0$. The sum rule

$$\sum_\lambda \omega^2(\mathbf{q},\lambda) = \omega_{pl}^2 \tag{7}$$

is valid for those frequencies. In the static case, expression (5) can be rewritten as

$$\frac{1}{\varepsilon(\mathbf{q},0)} = \sum_\lambda \left\{ (\mathbf{n}\mathbf{e}_{\mathbf{q}\lambda})^2 \left[ 1 - \frac{\omega_{pl}^2}{\omega^2(\mathbf{q},\lambda)} \right] \right\} \quad , \tag{8}$$

where $\mathbf{n} = \mathbf{q}/|q|$. Here, we took into account that

$$\sum_\lambda (\mathbf{n}\mathbf{e}_{\mathbf{q}\lambda})^2 = 1 \quad . \tag{9}$$

With the account for the sum rule (7), one can easily see that the right-hand side of expression (8) is always negative at any $\mathbf{q}$ vector. Notice that, according to expression (8), the inequality $\varepsilon(\mathbf{q}, 0) < 0$ is valid just in the stable phase of Wigner crystal, when all the quantities $\omega^2(\mathbf{q}, \lambda)$ are positively defined, i.e., $\omega^2(\mathbf{q}, \lambda) > 0$.

It follows from the above that in the Wigner crystal, the negative values of $\varepsilon(\mathbf{q}, 0)$ occur because the single mode of the plasma oscillations, that exists in gaseous and liquid states, is split to three modes in the solid state. This is due to the localization of charges and strong local-field effects in the crystal. Furthermore, as has been shown in Ref. 6, the negative values of $\varepsilon(\mathbf{q}, 0)$ occur in the classical one-component plasma when the interaction parameter $\Gamma = e^2/(aT)$ [here, $a = (4n/3)^{-1/3}$] is considerably smaller than its value at which the Wigner crystallization takes place, $\Gamma \approx 170$. As is shown in paper (6), the static dielectric function $\varepsilon(\mathbf{q}, 0)$ is negative virtually at all $\mathbf{q}$ vectors in the density range corresponding to $\Gamma > 40$. It is intriguing that at $\Gamma > 40$ the plasma oscillations have the negative dispersion [$\omega_{pl}^2(\mathbf{q}) = \omega_{pl}^2(0) - \alpha q^2$] and a finite linewidth. It is also shown in Ref. 6 that the negative values of $\varepsilon_{tot}(\mathbf{q}, 0)$ exist not only in model systems, but also in real systems, e.g., in melted table salt.

The above-considered systems possessing the negative static dielectric function $\varepsilon(\mathbf{q}, 0)$ have, of course, nothing to do with high-temperature superconductivity and with superconductivity in general. Within the 'folklore' approach, the negative values of $\varepsilon(\mathbf{q}, 0)$ ensure that the inequality $\lambda > \mu$ is fulfilled. The simple expression (1) can be applied as it is only for a hypothetical highly-compressed metal where the parameter of Coulomb interaction $r_s$ satisfies the condition $r_s \ll 1$. Here, $r_s = (3n/4\pi a_B^3)^{1/3}$ and $n$ and $a_B$ are the density and the Bohr radius of the electrons, respectively. In this case the dielectric function $\varepsilon_{tot}(\mathbf{q}, 0)$ can be written in the form [7, 8]

$$\frac{1}{\varepsilon_{tot}(\mathbf{q},0)} = \frac{4\pi e^2}{q^2 \varepsilon_{el}(q,0)} \sum_\lambda \left\{ (\mathbf{n}\mathbf{e}_{\mathbf{q}\lambda})^2 \left[ 1 - \frac{\omega_{jl}^2(\mathbf{q})}{\omega^2(\mathbf{q},\lambda)} \right] \right\} . \tag{10}$$

Here, $\varepsilon_{el}(\mathbf{q}, 0)$ is the static dielectric function of the electron gas, which can be expressed within RPA as

$$\varepsilon_{el}(q,0) = 1 + \frac{\chi^2}{q^2} , \tag{11}$$

$\omega(\mathbf{q}, \lambda)$ is the phonon frequency, and $\omega_{jl}^2(\mathbf{q})$ is the plasma frequency of the jelly-like model:

$$\omega_{jl}^2(q) = \frac{\omega_{pl}^2}{\varepsilon_{el}(q,0)} . \tag{12}$$

It is easy to show [7, 8] that for a stable phase of the highly-compressed metal the inequality

$$\omega_{jl}^2(q) \geq \omega^2(q,\lambda) \tag{13}$$

is valid. This inequality guarantees that in the highly-compressed metal the static dielectric function is negative at all $\mathbf{q}$ vectors and the inequality $\lambda - \mu^* > 0$ is satisfied. This means that, owing to the EPI, superconductivity can exist in this metal without any pseudization of the Coulomb repulsion. Certainly, at $r_s \ll 1$ the EPI coupling constant $\lambda$ is small and $T_c$ is low.

In our early work [7] we have calculated $\varepsilon_{tot}(\mathbf{q}, 0)$ for several simple metals (K, Al, Pb) and for hypothetical metallic hydrogen. We have shown [7] that for potassium, $\varepsilon_{tot}(\mathbf{q}, 0)$ is positive at all $\mathbf{q}$ vectors, while for Pb and metallic hydrogen it is negative at any $\mathbf{q}$. The reason for this is more or less the same as in the Wigner crystal. As follows from Eqs. (10) – (13), the phonon contribution to the dielectric function $\varepsilon_{tot}(\mathbf{q}, 0)$ is negative and exceeds the electron contribution because of the strong local-field effects in the system of localized ions.

Even for many conventional metals the electron-electron interaction that describes the superconducting state has a more complex form [8]. The main problem here is the successive calculation of the contribution of Coulomb interaction $\mu$ to the effective electron-electron interaction. As for the electron-phonon coupling constant $\lambda$, there are highly efficient density-functional methods (see, e.g., [9, 10]) for calculating $\lambda$, at least for conventional metals. Numerous calculations and tunneling measurements show that in many conventional metals and their compounds, $\lambda \gtrsim 1$. Moreover, as has been shown in Ref. 11, the coupling constant $\lambda$ in metallic hydrogen can reach the value of 6. In all these metals, the relation $\lambda \gtrsim 1$ is valid at average electron densities corresponding to $r_s \approx 1$. In a homogeneous electron gas at $r_s < 1$, the constant $\mu$ of the Coulomb electron-electron interaction can be written as [8]

$$\mu = \frac{r_s}{2\pi}.\tag{14}$$

This means that even at $r_s \approx 1$, the inequality $\mu < 1$ is still valid. We notice that the maximum value of $\mu$, as is said in Ref. 2, does not exceed $\mu = 1/2$. In ordinary metals, the Coulomb electron-electron interaction is only weakly affected by the crystal lattice. This is evident, for example, from the good coincidence of the plasma energy of electrons in a conventional metal and in a homogeneous electron gas with the same average density. In addition, the dispersion of plasma oscillations in the both systems is positive:

$$\omega_{\text{pl}}^2(\mathbf{q}) = \omega_{\text{pl}}^2(0) + \alpha q^2,\tag{15}$$

where $\alpha > 0$.

Contrary to the Coulomb electron-electron interaction, the EPI in crystals differs essentially from the case of simple homogeneous jelly-like models. In such models, both the electron-electron and electron-ion coupling are defined by the ordinary Coulomb interaction, hence $\lambda \approx \mu$. In the simple approximation, the EPI constant can be expressed as [8]

$$\lambda = \frac{3.01}{r_s}\overline{V}_{\text{ie}}^2(\mathbf{q})\left\langle\frac{\Omega_{\text{pl}}^2}{\omega^2}\right\rangle.\tag{16}$$

Here, $\overline{V}_{\text{ie}}^2(\mathbf{q})$ is the average square of the electron-ion pseudopotential, $\omega$ is the phonon frequency, and $\Omega_{\text{pl}}$ is the ion plasma frequency:

$$\Omega_{\text{pl}}^2 = \frac{4\pi N Z^2 e^2}{M}.\tag{17}$$

The average phonon frequencies in metals are considerably lower than the ion plasma frequency. This results in relatively high values of $\lambda$ as compared to $\mu$. As is noticed above, the reason for this is the correlations and local field effects in the ionic system, as well as an additional (relative to the Wigner crystal) screening of the frequencies of the ion vibrations by the conduction electrons. In many cases, in particular in alkali metals, the smallness of the average square of the electron-ion pseudopotential $\overline{V}_{\text{ie}}^2(\mathbf{q})$ is an essential factor that lowers the value of $\lambda$.

It follows from our analysis that the static dielectric constant not only can be negative, but must necessarily be negative in a number of systems, namely in their stable state. In some model metallic systems, for example, in highly-compressed metals with $r_s \ll 1$, the EPI constant $\lambda$ inevitably exceeds the constant $\mu$ of the Coulomb electron-electron interaction. There are, however, less rigorous, but quite plausible arguments [8, 12] in favor of the fact that in many conventional metals, $\lambda$ is actually larger than $\mu$ and the pseudization of the Coulomb contribution, i.e. the transition from $\mu$ to $\mu^* = \mu/[1 + \mu\ln(\varepsilon_F/\tilde{\omega})]$ is not important for the existence of the superconducting state.

Doubts are cast upon some other statements in paper [1], in particular, classifying interactions that exist in superconducting cuprates into mammoths, elephants, and mice. According to Ref. 1, the exchange interaction of the electrons at neighboring sites, as the strongest one, is appointed mammoth. As is mentioned in Ref. 1, this interaction occurs perturbatively in $1/U$, where $U$ is the repulsion of the electrons at one site. Thus, the exchange coupling constant is $J \sim t^2/U$, where $t$ is the overlap integral. We should take into account that $J$ describes not the attractive interaction of the electrons at neighboring sites, but the attraction of their spins only. The latter attraction is sufficient for antiferromagnetic ordering of the spins. In the framework of the simple Hubbard model, only with the repulsion of electrons at one site, the attractive exchange interaction is sufficient for the superconducting state to occur. In reality, however, the Coulomb repulsion of electrons exists at neighboring sites, too. As the ARPES experiments demonstrate [13], the holes in $CuO_2$-planes, at least in the optimally doped samples, form a system of strongly interacting Fermi particles. Most likely this system cannot be described as a Fermi-gas of weakly interacting quasi-particles. As is well known, in the system of strongly interacting Coulomb particles the interaction of holes at the average distance between them is of the order of their kinetic energy. This means that the Coulomb repulsion of the holes at neighboring sites is $V \sim t$, i.e., it is substantially stronger than the exchange interaction of the spins. Certainly, there are no reasons why $V$ might be much less than $t$. We do not know any publication on this subject, where such reasons are indicated. Likewise, there are no reasons for the EPI in cuprates to be much less than the exchange interaction or direct Coulomb repulsion at neighboring sites. The available first-principle calculations of the EPI in cuprates give rather contradictive results concerning the value of the EPI constant, but they do not indicate that the EPI is smaller than the exchange interaction [12].

Doubts are also cast upon the attempt of P.W. Anderson to relate the low-energy peculiarities of high-$T_c$ materials to the strong electron scattering caused by the large Hubbard repulsion $U$. The problem at issue is the existence of so-called 'kinks' in the single-particle excitation spectra of cuprates at 0.03–0.09 eV. Usually these peculiarities are ascribed to the electron coupling with bosonic modes. The physical nature of these bosonic modes has long been under discussion. Two possible candidates for these bosons are usually discussed, namely the phonons and the spin fluctuations. Sometimes this discussion looks curiously. Recently, two papers [14, 15] have been published concerning the calculation of the constant of electron coupling with the spin fluctuations. In Refs. 14 and 15, the same $t$–$J$ Hamiltonian and the same experimental data are used, but the obtained coupling constants differ by three orders of magnitude. More recently, peculiarities in the cuprates electron spectra at 0.3–0.5 eV have been also observed [16]. These high-energy anomalies are likely due to the Hubbard $U$ and to the exchange-correlation effects.

To conclude, it is necessary to explore in more detail what the "high-$T_c$ refrigerator" contains in terms of mammoths, elephants, and mice. We believe that there are still many other unsolved problems in high-$T_c$ materials.

The authors acknowledge fruitful discussions with many colleagues, first of all, with V.L. Ginzburg, M.R. Trunin, M.V. Sadovskii, and M. Kulič.